\documentclass[12pt]{article}
\usepackage{amsmath}
\usepackage{amssymb}
\usepackage{graphicx}
\usepackage{float}
\usepackage{indentfirst}
\usepackage{mathrsfs}

\usepackage{setspace}

\usepackage[labelfont=bf,labelsep=period]{caption}

\usepackage[numbers,sort&compress]{natbib}

\newtheorem{theorem}{Theorem}

\newtheorem{corollary}{Corollary}

\title{Quantum State Concentration and Classification of Multipartite Entanglement}

\author
{S. M. Zangi$^1$, Jun-Li Li$^{1,3}$, and Cong-Feng Qiao$^{1,2,3\ast}$\\ [0.2cm]
\normalsize{$^1$Department of Physics, University of Chinese Academy of Sciences,}\\
\normalsize{YuQuan Road 19A, Beijing 100049, China}\\[2pt]
\normalsize{$^2$ Department of Physics \& Astronomy, York University, Toronto, ON M3J 1P3, Canada}\\[2pt]
\normalsize{$^3$Key Laboratory of Vacuum Physics, University of Chinese Academy of Sciences}\\[3mm]
\normalsize{$^\ast$ To whom correspondence should be addressed; E-mail: qiaocf@ucas.ac.cn.}
}

\date{}

\begin{document}
\baselineskip24pt \maketitle
\begin{abstract}\doublespacing
Entanglement is a unique nature of quantum theory and has tremendous potential for application. Nevertheless, the complexity of quantum entanglement grows exponentially with an increase in the number of entangled particles. Here we introduce a quantum state concentration scheme which decomposes the multipartite entangled state into a set of bipartite and tripartite entangled states. It is shown that the complexity of the entanglement induced by the large number of particles is transformed into the high dimensions of bipartite and tripartite entangled states for pure quantum systems. The results not only simplify the tedious work of verifying the (in)equivalence of multipartite entangled states, but also are instructive to the quantum many-body problem involving multipartite entanglement.
\end{abstract}

\newpage

\section{Introduction}

Entanglement is regarded as an essential physical resource of quantum information sciences, which are responsible for the so called `second quantum revolution' \cite{Second-Rev}. Besides the developments of quantum algorithm \cite{Algorithms-Role, Algorithms-Multi} and quantum computation \cite{Rev-Measurement-Based}, every study related to many-body quantum system \cite{Entangled-Many-Body} would benefit from a deeper understanding of multipartite entanglement. The entanglement may be classified based on the different tasks it performs in quantum information processing, which forms the basis of the qualitative and quantitative characterizations of multipartite entanglement \cite{Two-Three}. Though an enormous amount of work in the literature has been dedicated to this subject \cite{Quantum-Entanglement, Entanglement-Detection}, a very limited information about the multipartite entanglement has been obtained. This is because the complexity of characterizing entanglement using classical parameters, i.e. coefficients of the quantum state in decomposition bases, grow dramatically with the particles and dimensions.

Two superficially different entangled states may be used to implement the same quantum information task identically if they are equivalent under local unitary (LU) and with different performances if they are equivalent under invertible local operators (stochastic local operations and classical communication, SLOCC). The LU equivalence of arbitrary multipartite entangled states could be understood via the high order singular value decomposition (HOSVD) \cite{LU-HOSVD, LU-HOSVD-2}, and an alternative method also exists for multi-qubit states \cite{LU-qubit, LU-qubit-e}. However, only the states with specific symmetries were explored by effective methods under SLOCC \cite{RG-states, N-symmetric}. While the coefficient matrix method is a practical but rather coarse grained classification method for multipartite entanglement \cite{N-Coefficient-M}, invariant polynomials encountered in distinguishing the inequivalent classes under SLOCC usually involve cumbersome rational expressions \cite{Infinite-SLOCC}. A recent study shows that the four-partite entanglement may be well understood through its sub-system's entanglement \cite{Four-SLOCC}. Then one may naturally ask whether the general multipartite entanglement could also be understood by the entangled subsystems, rather than by the classical parameters (coefficients of the quantum state) alone.

In this paper, we suggest a splitting scheme for the study of multipartite entanglement, which generalize the method of \cite{Four-SLOCC} to arbitrary multipartite states. By introducing virtual particles and performing a sequential of high order singular value decompositions, a multipartite entangled pure state is transformed into a set of states with only bipartite and tripartite entangled states. This set of states, which we call the core entangled states, forms a hierarchy structure. The concentration of multipartite entanglement to the core entangled states exhibits a similar structure as that of the tree tensor network state \cite{TTN-Vidal}. By applying to entanglement classification we find that two multipartite states are equivalent under LU or SLOCC if and only if their core entangled states in each hierarchy are equivalent under LU or SLOCC, respectively.

\section{Quantum state concentration}

An arbitrary $I_1\times I_2\times \cdots \times I_N$ dimensional multipartite quantum state has the form
\begin{eqnarray}
|\Psi\rangle = \sum_{i_1,i_2,\cdots, i_N=1}^{I_1,I_2,\cdots, I_N} \psi_{i_1i_2\cdots i_N} |i_1\rangle |i_2\rangle\cdots |i_N\rangle \; ,
\end{eqnarray}
where the complex numbers $\psi_{i_1i_2\cdots i_N} \in \mathbb{C}$ are coefficients of the state in the orthonormal bases $\{|i_1\rangle, |i_2\rangle, \cdots, |i_N\rangle \}$. In this form, the quantum state may be regarded as a high order tensor $\Psi$ whose tensor elements are $\psi_{i_1i_2\cdots i_N}$ and the inner product of two states of the same quantum system is defined as $\langle \Psi'|\Psi\rangle = \langle \psi'_{i_1\cdots i_N}|\psi_{i_1\cdots i_N} \rangle \equiv \sum_{i_1,i_2, \cdots, i_N=1}^{I_1, I_2, \cdots, I_N} \psi'^*_{i_1i_2\cdots i_N}\psi_{i_1i_2\cdots i_N}$. We group every two particles into a composite one, i.e., $(i_1i_2)(i_3i_4)\cdots (i_{N-1}i_N)$, and make the map $(i_{2k-1}i_{2k})\mapsto j_k$ such that $j_k = (i_{2k-1}-1)I_{2k}+i_{2k}$ (we may set $j_{(N+1)/2} = i_{N}$ for $N$ being odd). This rescaling of the quantum state can be expressed as
\begin{align}
|\Psi\rangle & =  \sum_{i_1,i_2,\cdots, i_N=1}^{I_1,I_2,\cdots,I_N}  \psi_{(i_1i_2)(i_3i_4)\cdots (i_{N-1}i_N)} |i_1i_2\rangle |i_3i_4\rangle \cdots |i_{N-1}i_N\rangle \nonumber \\
&= \sum_{j_1,j_2,\cdots, j_M=1}^{J_1,J_2,\cdots,J_M} \psi_{j_1j_2\cdots j_M} |j_1\rangle |j_2\rangle\cdots |j_M\rangle \; . \label{Rescale-One}
\end{align}
Now $\Psi$ may be regarded as an $M$-partite quantum state rescaled from the $N$-partite state.

For an $M$-order tensor $\Psi$ with dimensions of $J_1\times J_2\times \cdots \times J_M$, its $k$th mode matrix unfolding is represented by $\Psi_{(k)}$ which is a $J_k \times (J_{k+1} \cdots J_MJ_1J_2 \cdots J_{k-1})$ dimensional matrix with matrix elements $\psi_{j_k(j_{k+1} \cdots j_Mj_1j_2\cdots j_{k-1})}$ \cite{HOSVD}. The HOSVD of the $M$-partite state $\Psi$ is
\begin{eqnarray}
\Psi = U^{(1)}\otimes U^{(2)}\otimes \cdots \otimes U^{(M)}\, \Omega \; , \label{Formal-HOSVD}
\end{eqnarray}
where unitary matrices $U^{(k)} = (\vec{u}^{\,(k)}_1,\cdots, \vec{u}^{\,(k)}_{J_k})$ are composed of the left singular vectors of $\Psi_{(k)}$, and $\Omega$ is called the core tensor of $\Psi$ \cite{LU-HOSVD, HOSVD}. The core tensor $\Omega$ has the tensor elements $\omega_{j_1j_2\cdots j_M}$ and is all-orthogonal, i.e. $\langle \omega_{j_1\cdots j_k=\alpha\cdots j_M}|\omega_{j_1\cdots j_k=\beta\cdots j_M} \rangle = \delta_{\alpha\beta}$, $k \in \{1, \cdots, M\}$. Equation (\ref{Formal-HOSVD}) can also be written in form of tensor elements
\begin{eqnarray}
\Psi = \sum_{j_1,j_2,\cdots, j_M = 1}^{r_1,r_2,\cdots, r_M} \omega_{j_1j_2\cdots j_M}\, \vec{u}^{\,(1)}_{j_1} \circ \vec{u}^{\,(2)}_{j_2} \circ \cdots \circ \vec{u}^{\,(M)}_{j_M} \; . \label{HOSVD-Vector}
\end{eqnarray}
Here $r_k$ is the local rank of the $k$th mode matrix unfolding of $\Omega$, and $\vec{u}^{\,(k)}_{j_k}$ are $I_{2k-1} \times I_{2k}$ dimensional orthonormal vectors for $j_k \in \{1,\cdots, r_k\}$ with $\circ$ being the direct product. The singular vectors in the unitary matrix $U^{(k)}$ can be grouped into two parts according to the rank $r_k$,
\begin{equation}
U^{(k)} = (U^{(k)}_1,U^{(k)}_0)\; , \mathrm{where} \; U^{(k)}_1 \equiv (\vec{u}_1^{(k)}, \cdots, \vec{u}_{r_k}^{(k)})\; , \; U^{(k)}_0 \equiv (\vec{u}_{r_k+1}^{(k)}, \cdots, \vec{u}_{J_k}^{(k)}) \; . \label{group-singularvalues}
\end{equation}

We define the wrapping of an $(I_1 \times I_2)$-dimensional vector $\vec{u}$ into an $I_1 \times I_2$ matrix as follows \cite{Four-SLOCC}
\begin{eqnarray}
\mathcal{W}(\vec{u}) \equiv
\begin{pmatrix}
u_{1} & u_{I_1+1} & \cdots & u_{(I_2-1)I_1+1} \\
u_{2} & u_{I_1+2} & \cdots & u_{(I_2-1)I_1+2} \\
\vdots & \vdots & \ddots & \vdots \\
u_{I_1} & u_{2I_1} &\cdots & u_{I_2I_1}
\end{pmatrix}\; , \label{wrapping}
\end{eqnarray}
and the vectorization of a matrix is defined as $\mathcal{V}[\mathcal{W}(\vec{u})] \equiv \vec{u}$. An $(r\times I_1 \times I_2)$-dimensional tripartite pure state can be expressed in tuples of matrices, that is, $\{\Gamma_1,\cdots, \Gamma_{r}\}$ where $\Gamma_i \in \mathbb{C}^{I_1\times I_2}$ \cite{Four-SLOCC, LNN}. Hence, by wrapping the $(I_{2k-1} \times I_{2k})$-dimensional vector $\vec{u}_{j_k}^{(k)}$ into  an $I_{2k-1} \times I_{2k}$ matrix, we get a $r_{k} \times I_{2k-1}\times I_{2k}$ tripartite state $\Psi_{u_k}$ and its complementary state $\overline{\Psi}_{u_k}$ from the unitary matrix $U^{(k)} = (\vec{u}_1^{(k)}, \cdots, \vec{u}_{r_k}^{(k)}, \vec{u}_{r_k+1}^{(k)}, \cdots, \vec{u}_{J_k}^{(k)})$, i.e.
\begin{equation}
\Psi_{u_k} \equiv (\mathcal{W}(\vec{u}_{1}^{(k)}), \cdots, \mathcal{W}(\vec{u}_{r_k}^{(k)}))\; , \; \overline{\Psi}_{u_k} \equiv (\mathcal{W}(\vec{u}_{r_k+1}^{(k)}),\cdots, \mathcal{W}(\vec{u}_{J_k}^{(k)})) \; , \label{Tri-its-Comp}
\end{equation}
where $\mathcal{W}(\vec{u}^{(k)}_{i}) \in \mathbb{C}^{I_{2k-1} \times I_{2k}}$ are $(I_{2k-1} \times I_{2k})$-dimensional complex matrices \cite{Four-SLOCC}.

\begin{figure}\centering
\scalebox{0.55}{\includegraphics{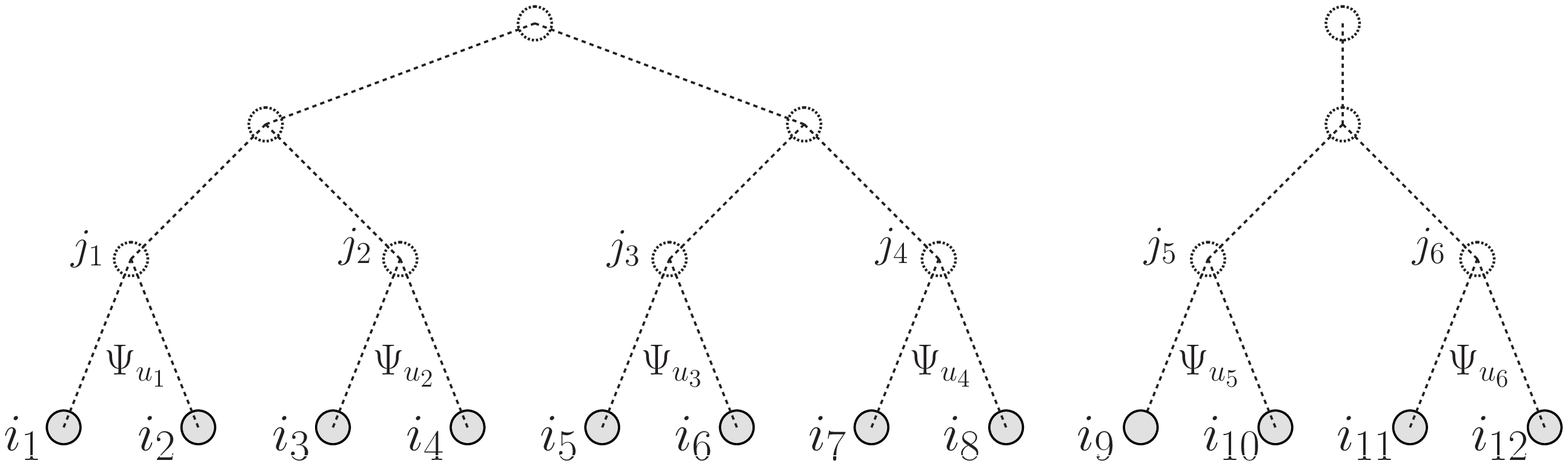}}
\caption{A 12-partite entangled state is first transformed into 6 tripartite states and one 6-partite state: $(\Psi_{u_1},\Psi_{u_2},\cdots, \Psi_{u_6}, \Omega_{j_1j_2\cdots j_6})$. Further rescaling may turn the 12-partite state into 10 tripartite states and one bipartite state.} \label{Fig-1}
\end{figure}

The $N$-partite state $\Psi$ now is rescaled and decomposed into $M$ tripartite states and one $M$-partite state
\begin{eqnarray}
\Psi = (\Psi_{u_1},\Psi_{u_2},\cdots, \Psi_{u_M}, \Omega_r)\; .
\end{eqnarray}
Here $\Psi_{u_k}$ are $r_k \times I_{2k-1} \times I_{2k}$ dimensional tripartite states as defined in equation (\ref{Tri-its-Comp}) and $\Omega_r$ is an $(r_1\times r_2 \times \cdots \times r_M)$-dimensional $M$-partite state whose coefficients are $\omega_{j_1j_2 \cdots j_M}$. Further rescaling of $\Omega_r$ as that of equation (\ref{Rescale-One}) and then wrapping the singular vectors would lead to another set of tripartite states where one may get a hierarchy structure of tripartite states for the multipartite entangled state in the end (see Fig. \ref{Fig-1}). We call this decomposition of a multipartite entangled state into bipartite and tripartite entangled states the quantum state concentration, which exhibits a form quite similar to the tree tensor network state \cite{TTN-Vidal}. To exemplify the application of the scheme in the quantum many-body problem, we apply the quantum state concentration technique to the multipartite entanglement classification.

For the equivalence (under LU operation or SLOCC) of two $N$-partite states $\Psi'$ and $\Psi$, we have the following theorem, which is a multipartite generalization of Ref. \cite{Four-SLOCC}.
\begin{theorem}
Two $N$-partite entangled states $\Psi'$ and $\Psi$ are equivalent if and only if the quantum states in the decompositions $\Psi'=(\Psi_{u_1'},\cdots,\Psi_{u'_M},\Omega'_r)$ and $\Psi = (\Psi_{u_1},\cdots,\Psi_{u_M}, \Omega_r)$ are equivalent in the following way:
\begin{align}
\Psi_{u'_{k}} & = P^{(k)} \otimes A_{2k-1}\otimes A_{2k}\, \Psi_{u_k}\;, \forall\, k\in \{1,\cdots, M\} \; , \label{Tri-equation}\\
\Omega_r & = P^{(1)}\otimes \cdots \otimes P^{(M)} \, \Omega'_r \; , \label{M-Tensor}
\end{align}
where $A_{i}$, $P^{(k)}$ are all invertible (or unitary) matrices for SLOCC (or LU) equivalence. \label{Theorem-1}
\end{theorem}

\noindent {\bf Proof:} First, if $\Psi' = A_1\otimes \cdots \otimes A_N \,\Psi$, then the HOSVD of the rescaled $M$-partite $\Psi'$ and $\Psi$ is equivalent to
\begin{eqnarray}
U'^{(1)} \otimes \cdots \otimes  U'^{(M)} \, \Omega'  = (A_1\otimes A_2) U^{(1)} \otimes \cdots \otimes (A_{2M-1}\otimes A_{2M}) U^{(M)} \, \Omega\; . \label{N-hosvd-omega}
\end{eqnarray}
Here $\Omega'$ and $\Omega$ have the nonzero parts of $\Omega_r'$ and $\Omega_r$ respectively. Substituting the QR factorization $(A_{2k-1}\otimes A_{2k}) U^{(k)} = Q_{u_k}R_{u_k}$ into equation (\ref{N-hosvd-omega}), and applying HOSVD to $R_{u_1}\otimes \cdots \otimes R_{u_M} \Omega$, we get
\begin{eqnarray}
R_{u_1}\otimes \cdots \otimes R_{u_M} \Omega & = & X_{u_1}\otimes \cdots \otimes X_{u_M} \Omega' \; , \label{th-equa-RXO} \\
U'^{(k)} & = & Q_{u_k}X_{u_k}\; , \forall\, k \in \{1,\cdots, M\} \; ,\label{th-equa-uqx}
\end{eqnarray}
where $X_{u_k}$ are unitary matrices. Equations (\ref{th-equa-RXO}, \ref{th-equa-uqx}) also give
\begin{eqnarray}
\Omega & = & (R_{u_1}^{-1}X_{u_1}) \otimes \cdots \otimes (R_{u_M}^{-1} X_{u_M})\, \Omega' \; , \label{ptilde-form} \\
U'^{(k)} & = & (A_{2k-1} \otimes A_{2k}) U^{(k)} (R_{u_k}^{-1} X_{u_k})  \; , \forall\, k \in \{1,\cdots, M\}\; . \label{ptilde-unitary}
\end{eqnarray}
Let $\widetilde{P}^{(k)} \equiv R_{u_k}^{-1}X_{u_k}$, because $\Omega'$ and $\Omega$ have the same local ranks of $r_k$, equation (\ref{ptilde-form}) leads to
\begin{eqnarray}
\widetilde{P}^{(k)} =
\begin{pmatrix}
  P^{(k)} & Y^{(k)} \\
  0 & \overline{P}^{(k)}
\end{pmatrix}\; . \label{Ptilde-k}
\end{eqnarray}
Here $P^{(k)} \in \mathbb{C}^{r_k\times r_k}$ and $\overline{P}^{(k)} \in \mathbb{C}^{(I_{2k-1}\cdot I_{2k}-r_k)\times (I_{2k-1}\cdot I_{2k}-r_k)}$ are invertible matrices, and $\widetilde{P}^{(k)}$ are unitary if all matrices $A_{j}$ are unitary which is easy to see from equation (\ref{ptilde-unitary}). As the tensor elements $\omega'_{j_1j_2\cdots j_M}$ and $\omega_{j_1j_2\cdots j_M}$ of the core tensors $\Omega$ and $\Omega'$ are nonzero only for $1\leq j_{k}\leq r_k$, $\forall k \in \{1,\cdots, M\}$, equations (\ref{ptilde-form}) and (\ref{Ptilde-k}) lead to equation (\ref{M-Tensor}). Taking equation (\ref{Ptilde-k}) into equation (\ref{ptilde-unitary}), we have
\begin{align}
(\vec{u}\,'^{(k)}_{\!1}, \vec{u}\,'^{(k)}_{\!2},\cdots, \vec{u}\,'^{(k)}_{\!r_k}) = A_{2k-1}\otimes A_{k} (\vec{u}^{(k)}_1,\vec{u}^{(k)}_2,\cdots, \vec{u}^{(k)}_{r_k}) P^{(k)} \; , \label{partioned-uu}
\end{align}
where $\vec{u}\,'^{(k)}_{\!i}$ and $\vec{u}^{(k)}_i$ are from $U'^{(k)} = (U'^{(k)}_1, U'^{(k)}_0)$ and $U^{(k)} = (U^{(k)}_1, U^{(k)}_0)$ based on the definition in equation (\ref{group-singularvalues}). The wrapping operations make $(\mathcal{W}( \vec{u}\,'^{(k)}_{\!1}), \cdots, \mathcal{W}(\vec{u}\,'^{(k)}_{\!r_k})) = \Psi_{u_k'}$ and $(\mathcal{W}( \vec{u}^{(k)}_{1}), \cdots, \mathcal{W}(\vec{u}^{(k)}_{r_k})) = \Psi_{u_k}$, therefore equation (\ref{partioned-uu}) is equivalent to equation (\ref{Tri-equation}).

Second, equation (\ref{Tri-equation}) may be expressed in form of
\begin{eqnarray}
U_1'^{(k)} = (A_{2k-1}\otimes A_{2k}) U_1^{(k)} P^{(k)} \; , \forall\, k\in \{1,\cdots, M\} \; ,
\end{eqnarray}
where $U_1'^{(k)}=(\vec{u}\,'^{(k)}_{\!1},\cdots, \vec{u}\,'^{(k)}_{\!r_k})$ and $U_1^{(k)}=(\vec{u}^{(k)}_{1},\cdots, \vec{u}^{(k)}_{r_k})$ are from $U'^{(k)} = (U'^{(k)}_1, U'^{(k)}_0)$ and $U^{(k)} = (U_1^{(k)}, U_0^{(k)})$. We are legitimate to construct the matrix $\widetilde{P}^{(k)} = \begin{pmatrix}
P^{(k)} & Y^{(k)} \\
0 & \overline{P}^{(k)}
\end{pmatrix}$ such that
\begin{eqnarray}
U'^{(k)} = (A_{2k-1}\otimes A_{2k}) U^{(k)}
\begin{pmatrix}
P^{(k)} & Y^{(k)} \\
0 & \overline{P}^{(k)}
\end{pmatrix}\; , \label{construct-UU}
\end{eqnarray}
where $\overline{P}^{(k)}$ is invertible (unitary when $A_j$ are unitary). The decomposition $\Psi'=(\Psi_{u_1'},\cdots,\Psi_{u'_M},\Omega'_r)$ can be expressed as the follow
\begin{eqnarray}
\Psi' & = & U'^{(1)} \otimes \cdots \otimes U'^{(M)} \Omega' \nonumber \\
& = & (A_{1}\otimes A_{2}) U^{(1)}  \widetilde{P}^{(1)}\otimes \cdots \otimes (A_{2M-1}\otimes A_{2M}) U^{(M)} \widetilde{P}^{(M)} \Omega' \nonumber \\
& = & (A_{1}\otimes A_{2}) U^{(1)} \otimes \cdots \otimes (A_{2M-1}\otimes A_{2M}) U^{(M)} \Omega \nonumber \\
&= & A_1\otimes A_2 \otimes \cdots \otimes A_{2M} \, \Psi \; .
\end{eqnarray}
Here, equation (\ref{construct-UU}) is used in the second equality and equation (\ref{M-Tensor}) is used in the third equality. Therefore, $\Psi'$ and $\Psi$ are equivalent under SLOCC or LU when $A_i$ are invertible or unitary. Q.E.D.

The Theorem \ref{Theorem-1} decomposes the $N$-partite entangled state into $M$ tripartite states and one $M$-partite state, where $M = \lceil \frac{N}{2}\rceil$ is the smallest integer greater than or equal to $N/2$. The $M$-partite state could be further rescaled and turn into another set of $\lceil \frac{M}{2}\rceil$ tripartite states and one $\lceil \frac{M}{2}\rceil$-partite entangled state. Along this line, one may finally get a hierarchy of tripartite entangled states and one bipartite entangled state (see Fig. \ref{Fig-1}). This scheme therefore reduces the entanglement classifications of multipartite state to that of only bipartite and tripartite states, and makes the tripartite entanglement a key ingredient of quantum entanglement.

The fact that the set of tripartite and bipartite entangled states represents faithfully the multipartite entanglement can be understood as follows. The number of parameters needed to characterize the entanglement classes under SLOCC for $I_1\times I_2 \times \cdots \times I_N$ quantum state is \cite{Two-Three}
\begin{equation}
\mathcal{N}_{I_1\times \cdots \times I_N} = 2(I_1I_2\cdots I_N-1) - 2\sum_{i=1}^N (I_i^2-1) \; .
\end{equation}
In the decomposition $\Psi = (\Psi_{u_1}, \cdots, \Psi_{u_M}, \Omega_r)$ of Theorem \ref{Theorem-1}, the number of parameters becomes $\mathcal{N}_3 + \mathcal{N}_{M}$, where
\begin{align}
\mathcal{N}_{3} & = \sum_{k=1}^M  \left[2(r_kI_{2k-1}I_{2k} -1) - 2(I_{2k-1}^2+I_{2k}^2 - 2) \right] \; , \\ \mathcal{N}_{M} & =
2(r_1r_2\cdots r_M -1) -2 \sum_{k=1}^M(r_k^2-1) \; .
\end{align}
Here $2(I_{2k-1}^2 + I_{2k}^2-2)$ are induced by $A_{2k-1}$ and $A_{2k}$ in the $M$ tripartite entangled states and $2\sum_{k=1}^M (r_k^2-1)$ are induced by $P^{(k)}$ in the $M$-partite entangled states, according to equations (\ref{Tri-equation}) and (\ref{M-Tensor}) in Theorem \ref{Theorem-1}. The number $\mathcal{N}_{3}+ \mathcal{N}_{M}$ equals $\mathcal{N}_{I_1\times \cdots \times I_N}$ in the worst case of $r_k = I_{2k-1}I_{2k}$ in the rescaling process. Along this line, we will finally get a set of states with bipartite and tripartite entangled states only and the complexity of characterizing the entanglement of multipartite is transformed into the large numbers and high dimensions of the tripartite and bipartite entangled states in the set.

To illustrate how do the parameters in the multipartite state transform under the decomposition of theorem \ref{Theorem-1}, we present explicit examples of a four-qubit and a six-qubit states. Considering the four-qubit state $|\Psi\rangle = a_1|0001\rangle + a_2|0010\rangle + a_3|0100\rangle + a_4|1000\rangle$, where we assume $a_i \in \mathbb{R}$ for the sake of illustration, the state contains three independent real parameters (four parameters with one normalization constraint). The four particles may be grouped as
\begin{align}
|\Psi\rangle & = a_1|(00)(01)\rangle + a_2 |(00)(10)\rangle + a_3 |(01)(00)\rangle + a_4|(10)(00)\rangle \nonumber \\
& = a_1|01\rangle + a_2|02\rangle + a_3|10\rangle + a_4|20\rangle \; . \label{Example-4}
\end{align}
The last line in equation (\ref{Example-4}) is a bipartite state of $4\times 4$, and can be represented by a matrix whose the singular value decomposition is
\begin{align}
\Psi & = \begin{pmatrix}
0 & a_1 & a_2 & 0 \\
a_3 & 0 & 0 & 0 \\
a_4 & 0 & 0 & 0 \\
0 & 0 & 0 & 0
\end{pmatrix} = U\Lambda V^{\dag} =
\begin{pmatrix}
1 & 0 & 0 & 0 \\
0 & \frac{a_3}{\sqrt{a_3^2+a_4^2}} & 0 & \frac{-a_4}{\sqrt{a_3^2+a_4^2}} \\
0 & \frac{a_4}{\sqrt{a_3^2+a_4^2}} & 0 & \frac{a_3}{\sqrt{a_3^2+a_4^2}} \\
0 & 0 & 1 & 0
\end{pmatrix} \nonumber \\
& \hspace{1cm} \cdot \begin{pmatrix}
\sqrt{a_1^2+a_2^2} & 0 & 0 & 0 \\
0 & \sqrt{a_3^2+a_4^2} & 0 & 0 \\
0 & 0 & 0 & 0 \\
0 & 0 & 0 & 0
\end{pmatrix} \cdot
\begin{pmatrix}
0 & 1 & 0 & 0 \\
\frac{a_1}{\sqrt{a_1^2+a_2^2}} & 0 & 0 & \frac{-a_2}{\sqrt{a_1^2+a_2^2}} \\
\frac{a_2}{\sqrt{a_1^2+a_2^2}} & 0 & 0 & \frac{a_1}{\sqrt{a_1^2+a_2^2}} \\
0 & 0 & 1 & 0
\end{pmatrix}^{\dag} \; .
\end{align}
Based on equation (\ref{Tri-its-Comp}), we obtained one bipartite state $\psi_{\Lambda} = \mathrm{diag}\{\sqrt{a_1^2+a_2^2}, \sqrt{a_3^2+a_4^2}\}$, and two tripartite states
\begin{align}
\psi_{u} = \{\begin{pmatrix}
1 & 0 \\
0 & 0
\end{pmatrix}, \frac{1}{\sqrt{a_3^2+a_4^2}}\begin{pmatrix}
0 & a_4 \\
a_3 & 0
\end{pmatrix} \}\; , \; \psi_{v} = \{ \frac{1}{\sqrt{a_1^2+a_2^2}}\begin{pmatrix}
0 & a_2 \\
a_1 & 0
\end{pmatrix}, \begin{pmatrix}
1 & 0 \\
0 & 0
\end{pmatrix} \} \; ,
\end{align}
where there is one free parameter in each of them (note that $\frac{a_3^2}{a_3^2+a_4^2}+ \frac{a_4^2}{a_3^2+a_4^2}=1$). In this example, the parameters in the multipartite entangled state $|\psi\rangle$ are evenly distribute among the decomposed tripartite and bipartite entangled states. As the number of core entangled states grows, there will be fewer parameters in each individual decomposed state, which results in a simplification to the practical entanglement classification.

Considering the six-qubit quantum state $|\Phi\rangle = b_1|000000\rangle + b_2 |010101\rangle + b_3 |101010\rangle + b_4|111111\rangle$ with $b_i \in \mathbb{R}$, we may group the six particles as follows
\begin{align}
|\Phi\rangle & = b_1|(00)(00)(00)\rangle + b_2 |(01)(01)(01)\rangle + b_3 |(10)(10)(10)\rangle + b_4|(11)(11)(11)\rangle \nonumber \\
& = b_1|000\rangle + b_2|111\rangle + b_3|222\rangle + b_4|333\rangle \; , \label{Example-61}
\end{align}
where the last line represents a tripartite state of $4\times 4\times 4$. An HOSVD to this tripartite state leads to
\begin{align}
\phi_{u_k} & = \{
\begin{pmatrix}
1 & 0 \\
0 & 0
\end{pmatrix}, \begin{pmatrix}
0 & 1 \\
0 & 0
\end{pmatrix},
\begin{pmatrix}
0 & 0 \\
1 & 0
\end{pmatrix},
\begin{pmatrix}
0 & 0 \\
0 & 1
\end{pmatrix} \}\; , \; k\in \{1,2,3\} \; , \\
|\phi_{\Omega}\rangle & =  b_1|000\rangle + b_2|111\rangle + b_3|222\rangle + b_4|333\rangle \; . \label{state-444}
\end{align}
That is, we get three $4\times 2\times 2$ entangled states, $\phi_{u_1}$, $\phi_{u_2}$, and $\phi_{u_3}$, and one $4\times 4\times 4$ state $|\phi_{\Omega}\rangle$. Further decomposition of equation (\ref{state-444}) may be performed according to the grouping of $|\phi_{\Omega} \rangle =  b_1|(00)0\rangle + b_2|(11)1\rangle + b_3|(22)2\rangle + b_4|(33)3\rangle$. However, we may stop at equation (\ref{state-444}), as we have already decomposed the multipartite state into only tripartite states. In this example, all the parameters in $|\Phi\rangle$ transform and concentrate into the high dimensional $4\times 4\times 4$ tripartite state, and there is no parameter in the other three tripartite entangled states $\phi_{u_k}$, $k\in \{1,2,3\}$.

These two explicit examples provide an understanding of how our method works. The parameters of the multipartite entangled state are redistributed and/or concentrated into the core entangled states, which are at most tripartite entangled. In the following we present two practical Corollaries for verifying the equivalence of tripartite entanglement under SLOCC and LU. The realignment of a matrix $A\in \mathbb{C}^{I_1\cdot I_2\times I_1\cdot I_2}$ according to the factorization of $I_1\times I_2$ is defined as \cite{Matrix-Realignment}
\begin{eqnarray}
\mathcal{R}(A) \equiv
\begin{pmatrix}
\mathcal{V}(A_{11}), \cdots, \mathcal{V}(A_{I_11}),
\mathcal{V}(A_{12}), \cdots, \mathcal{V}(A_{I_12}),\cdots,\mathcal{V}(A_{I_1I_1})
\end{pmatrix}^{\mathrm{T}} \; , \nonumber
\end{eqnarray}
where $\mathcal{R}(A) \in \mathbb{C}^{I_1\cdot I_1\times I_2\cdot I_2}$, and $A_{ij}\in \mathbb{C}^{I_2\times I_2}$ are the submatrices of $A$,
\begin{eqnarray}
A = \begin{pmatrix}
A_{11} & A_{12} & \cdots & A_{1I_1} \\
A_{21} & A_{22} & \cdots & A_{2I_1} \\
\vdots & \vdots & \ddots & \vdots   \\
A_{I_11} & A_{I_12} & \cdots & A_{I_1I_1}
\end{pmatrix} \; .
\end{eqnarray}
For two $r\times I_1\times I_2$ genuine tripartite entangled states $\Psi_{u'} = (\mathcal{W}(\vec{u}'_1), \cdots, \mathcal{W}(\vec{u}'_r))$ and $\Psi_{u} = (\mathcal{W}(\vec{u}_1), \cdots, \mathcal{W}(\vec{u}_r))$, we may construct their complementary states, i.e. $\overline{\Psi}_{u'} = (\mathcal{W}(\vec{u}'_{r+1}), \cdots, \mathcal{W}(\vec{u}'_{I_1\cdot I_2}))$ and $\overline{\Psi}_{u} = (\mathcal{W}(\vec{u}_{r+1}), \cdots, \mathcal{W}(\vec{u}_{I_1\cdot I_2}))$, where $\vec{u}'_i$ and $\vec{u}_i$ are $I_1\cdot I_2$ dimensional vectors, and $U' = (\vec{u}'_1,\cdots, \vec{u}'_{I_1\cdot I_2})$ and $U = (\vec{u}_1,\cdots, \vec{u}_{I_1\cdot I_2})$ are invertible matrices \cite{Four-SLOCC}. We have the following Corollaries.
\begin{corollary}
Two $r\times I_1\times I_2$ dimensional entangled quantum states $\Psi_{u'}$ and $\Psi_{u}$ are equivalent under local operators, i.e. $|\Psi_{u'}\rangle = P \otimes A_{1}\otimes A_{2}\, |\Psi_{u}\rangle$, if and only if there exist $\widetilde{P}= \begin{pmatrix}
P & Y \\
0 & \overline{P}
\end{pmatrix}\in \mathbb{C}^{I_1\cdot I_2\times I_1\cdot I_2}$ such that
\begin{eqnarray}
\mathrm{rank}[\mathcal{R}(U\widetilde{P} U'^{-1})] = 1\; . \label{coro-rank-eq}
\end{eqnarray}
Here $\mathcal{R}$ is the matrix realignment according to the factorization of $I_1\times I_2$; $\widetilde{P}$ and $U\widetilde{P} U'^{-1}$ are invertible (unitary) for SLOCC  (LU) equivalences. \label{corollary-1}
\end{corollary}

\noindent {\bf Proof:} It has been shown that, $\Psi_{u'}$ and $\Psi_{u}$ are equivalent under $P$, $A_1$, and $A_2$ if and only if \cite{Four-SLOCC}
\begin{eqnarray}
(U_1',U_0') = (A_{1} \otimes A_{2}) (U_1,U_0)
\begin{pmatrix}
P & Y \\
0 & \overline{P}
\end{pmatrix}\; . \label{UPAUP}
\end{eqnarray}
Therefore, $(A_1^{-1}\otimes A_2^{-1}) = U\widetilde{P}U'^{-1}$. According to Lemma 3 of Ref. \cite{2LMN}, $U\widetilde{P}U'^{-1}$ is direct product of two unitary or invertible matrices if and only if $U\widetilde{P} U'^{-1}$ is unitary or invertible and $\mathcal{R}(U\widetilde{P}U'^{-1})$ has rank 1.
Q.E.D.

\begin{corollary}
Two $r\times I_1\times I_2$ dimensional entangled quantum states $\Psi_{u'}$ and $\Psi_{u}$ are equivalent under local operators, i.e. $|\Psi_{u'}\rangle = P \otimes A_{1}\otimes A_{2}\, |\Psi_{u}\rangle$, if and only if there exist $\widetilde{P} = \begin{pmatrix}
P & Y \\
0 & \overline{P}
\end{pmatrix}\in \mathbb{C}^{I_1\cdot I_2\times I_1\cdot I_2}$ such that
\begin{eqnarray}
\forall \vec{a}\,,\, \mathcal{F}[\mathcal{W}(U\widetilde{P} U'^{-1} \vec{a})] = \mathcal{F}[\mathcal{W}(\vec{a})]\; .
\end{eqnarray}
Here, $\vec{a}$ is an arbitrary $I_1\cdot I_2$ dimensional vector; for SLOCC equivalence, $\mathcal{F}$ denotes the rank; for LU equivalence, $\widetilde{P}$ should be unitary and $\mathcal{F}$ denotes a concave, symmetric, and strictly increasing function on singular values of matrices with $\mathcal{F}(0)=0$. \label{corollary-2}
\end{corollary}

\noindent {\bf Proof:} The operator $\Phi = U\widetilde{P}U'^{-1}$ induces a linear map $\varphi: \mathbb{C}^{I_1\times I_2} \mapsto \mathbb{C}^{I_1\times I_2}$ for the wrapping $\mathcal{W}$: $\mathcal{W}(\Phi \vec{a}) = \varphi[\mathcal{W}(\vec{a})]$ \cite{Four-SLOCC}. The proof the Corollary can be carried out straightforwardly by the application of linear preserver problem with local ranks \cite{LPP-rank} and matrix norms \cite{LPP-unitary}. Q.E.D.

With the state concentration technique, the verification of SLOCC and LU equivalence of multipartite entanglement turns to the bipartite and tripartite entanglement classifications. The Corollaries \ref{corollary-1} and \ref{corollary-2} further simplify the verification of equivalent relations for tripartite entanglement. Note that the proposed method employs only linear equations in the verification procedure (see equation (\ref{coro-rank-eq})) and detailed information of the connecting matrices, i.e. $A_1,\cdots A_N$, is not the prerequisite for both SLOCC and LU equivalences of two tripartite entangled states \cite{Four-SLOCC}.

\section{Conclusion}

The characterization of multipartite entanglement is a longstanding tough issue in quantum information, due to the dramatic increase in the number of parameters characterizing it. In this work a quantum state concentration technique is introduced, which turns the multipartite entangled state into a set of bipartite and  tripartite entangled states, and the complexity of the entanglement characterization for multiple particles is transformed into that of large numbers and high dimensions of tripartite and bipartite entangled states in the set. By exploring the method, the classification of multipartite entanglement under SLOCC or LU is accomplished by classifying only the core entangled states, i.e. tripartite and bipartite entangled states. The results indicate that the multipartite entanglement is no more complex than the tripartite entangled states of high enough dimensions. Considering the implicit relation to the tree tensor network state, the scheme presented here may also be instructive in other studies concerning quantum multipartite states, e.g., condensed matter physics \cite{Entangled-Many-Body} and quantum chemistry \cite{TTNS-chem}.

\section*{Acknowledgements}

\noindent
This work was supported in part by the Ministry of Science and Technology of the Peoples' Republic of China(2015CB856703); by the Strategic Priority Research Program of the Chinese Academy of Sciences, Grant No.XDB23030100; and by the National Natural Science Foundation of China(NSFC) under the grants 11375200 and 11635009. S.M.Z. is also supported in part by the CAS-TWAS fellowship.

\end{document}